\documentclass[osajnl,twocolumn,showpacs,superscriptaddress,10pt]{revtex4-1} %% use 11pt for Applied Optics
\usepackage{amsmath,amssymb,graphicx}
\usepackage[percent]{overpic}
\usepackage[euler]{textgreek}
\usepackage{xcolor}
\begin{document}

\title{Broadband, unpolarized repumping and clearout light sources for Sr$^+$ single-ion clocks }

\author{T. Fordell} \email{Corresponding author: thomas.fordell@vtt.fi}
\author{T. Lindvall}
\affiliation{VTT Technical Research Centre of Finland Ltd, Centre for Metrology MIKES, P.O.Box 1000, FI-02044 VTT, Finland}
\author{P. Dub\'e}
\author{A.A. Madej}
\affiliation{Frequency and Time Group, Measurement Science and Standards Portfolio, National Research Council of Canada,
Ottawa, Canada K1A 0R6}
\author{A.E. Wallin}
\author{M. Merimaa}
\affiliation{VTT Technical Research Centre of Finland Ltd, Centre for Metrology MIKES, P.O.Box 1000, FI-02044 VTT, Finland}

\begin{abstract}
Future transportable optical clocks require compact and reliable light sources. Here, broadband, unpolarized repumper and state clearout sources for Sr$^+$ single-ion optical clocks are reported. These turn-key devices require no frequency stabilization nor external modulators. They are fiber based, inexpensive, and compact. Key characteristics for clock operation are presented, including optical spectra, induced light shifts and required extinction ratios. Tests with an operating single-ion standard  show a clearout efficiency of 100\%. Compared to a laser-based repumper, the achievable fluorescence rates for ion detection are a few tens of per cent lower. The resulting ion kinetic temperature is 1--1.5\;mK, near the Doppler limit of the ion system. Similar repumper light sources could be made for Ca$^+$ (866 nm) and Ba$^+$ (650\;nm) using semiconductor gain media. \\
\\

This paper was published in Optics Letters and is made available as an electronic reprint with the permission of OSA. The paper can be found at the following URL on the OSA website: http://www.opticsinfobase.org/ol/abstract.cfm?uri=ol-40-8-1822. Systematic or multiple reproduction or distribution to multiple locations via electronic or other means is prohibited and is subject to penalties under law.

\end{abstract}
%\ocis{(300.6520) Spectroscopy, trapped ion; (120.3940)  Metrology; (060.2310) Fiber optics;  (230.2285)   Fiber devices and optical amplifiers.}% REPLACE WITH CORRECT OCIS CODES FOR YOUR ARTICLE
                          % NOTE: \ocis{} IS ALIASED TO \pacs{} BUT MUST
                          % FORMAT THE TERMS CORRECTLY FOR EACH JOURNAL
\maketitle %% required
The fractional uncertainties of several types of optical clocks are at or approaching the 10$^{-18}$ level \cite{Chou2010a, Hinkley2013a, Bloom2014a, Dube2014a, Barwood2014a}, %which is two orders of magnitude below the best Cs fountain clocks 
 which is equivalent to a change in the gravitational redshift caused by a height difference of only 1\;cm. %With fiber optic techniques, remote optical clocks can be compared to high precision even with very short integration times \cite{Calosso2014a}. Nevertheless, 
%Therefore, clock comparisons at the lowest uncertainty level must be done locally, since the required level of knowledge on the gravitational potential is not available. , transportable optical clocks (TOCs) are needed. 
Since such detailed knowledge on the gravitational potential is currently not available, clock comparisons at the lowest uncertainty level must be done locally. Therefore, transportable optical clocks (TOCs) are needed. Besides timekeeping, TOCs will aid the search for variations in the fundamental constants \cite{Godun2014a}, and their gravitational sensitivity could find interesting applications such as geodesy \cite{Chou2010b}. %And together with the extreme short term stability of probe laser systems, 
 TOCs could also replace masers as local oscillators at remote radio telescopes and thereby improve resolution in very-long baseline interferometry.

%Before the SI second can be re-defined in terms of an optical transition, optical clocks constructed around the world must be compared at the lowest-possible uncertainty level. 
%Local comparisons are needed, which requires at least one of the clocks to be transportable. With fractional uncertainties of optical clocks approaching the 10$^{-18}$ level \cite{Chou2010a, Hinkley2013a, Bloom2014a, Dube2014a, Barwood2014a}, which corresponds to a height difference of only 1\;cm that is far below the current knowledge of the geoid, transportable optical clocks could also find very interesting applications in, e.g., geodesy \cite{Chou2010b}. VLBI? FINE STRUCTURE CONSTANT? 

Practical transportability calls for compactness and robustness on all parts of a clock. In addition to the extremely stable clock laser, an optical clock requires several other light sources that must address electronic transitions in the reference ion or atoms.
Many optical frequency/time references require 'repumper' light sources to provide optical pumping out of undesired metastable levels. In addition, rapid resetting of the upper state of a clock transition via a 'clearout' light source  is frequently required so that the optical reference can have low dead time and frequency instabilities approaching the quantum measurement limits of the system.  If lasers are used, wavelength stabilization is required, which adds complexity and calls for continuous operation of the lasers. % If the requirement of frequency stabilization can be relaxed, a considerable simplification of the light source ensemble results. 
\begin{figure}
{\includegraphics[width=0.75\columnwidth]{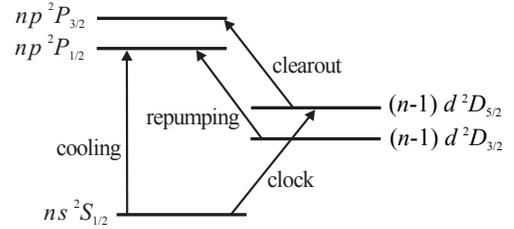}}
\caption{Partial energy level diagram of Sr$^+$ ($n=5$). Other clock ions with similar structure include Ca$^+$ \mbox{($n=4$)}, and Ba$^+$ ($n=6$).}% The diagram shows the 1092~nm P-D repumper transition and the 1033~nm transition that is used to depopulate the metastable $^2$D$_{5/2}$ state.}% after an excitation event on the reference/clock transition.}
\label{levelDiagram}
\end{figure}
Frequency stabilization is not needed if broadband sources are used, as proposed and theoretically studied in \cite{Lindvall2013a} for the case of the repumper and clearout light sources needed in single-ion optical clocks. If the broadband repumper is also unpolarized, the formation of dark states \cite{Lindvall2012a} is prevented in a zero magnetic field without any external modulators \cite{Barwood1998a, Berkeland1998b}. This article reports on the construction of such broadband, amplified spontaneous emission (ASE) sources for a Sr$^+$ single-ion optical clock. Key characteristics, including output spectra, achievable fluorescence rates, induced light shifts, modulation response and ion kinetic temperatures are discussed below. 

%Their suitability for use in optical clocks is verified by comparing their performance against laser based sources at the Sr$^+$ single-ion clock facility of the National Research Council of Canada \cite{Dube2013a, Dube2014a}.

%Broadband source at these wavelengths -> Gain media -> semiconductor yes -> superluminescent diodes commercially available, powerful and fiber coupled but power level too low.  (polarized and two orders of magnitude too low power for the repumper). Luckily ytterbium readily available -> cross-sections -> ytterbium works for both repumper and clear out.

%\section{Sr$^+$ Single Ion System}

The $^{88}$Sr$^+$ system (Fig.\;\ref{levelDiagram}), whose reference or clock line is the dipole forbidden 5s\,$^{2}$S$_{1/2}\rightarrow4$d\,$^{2}$D$_{5/2}$ transition at 445~THz (674 nm), is one of the single ion systems being studied as a potential optical atomic clock. The ion is laser cooled on the 422-nm  5s\,$^2$S$_{1/2}\rightarrow5$p\,$^2$P$_{1/2}$ resonance line transition using a  laser source. With a natural linewidth of 22~MHz, the ion kinetic temperature can reach a Doppler cooling limit of  0.5~mK. When probed on the 674~nm reference transition, interruption of the strong S-P fluorescence corresponds to probe excitation into the upper state of the clock transition and is thus used to sense the excitation with near unity detection efficiency. In addition to these two key laser-based sources, other light sources are necessary for the ion reference to operate effectively. Due to dipole-allowed relaxation from the 5p\,$^2$P$_{1/2}$ level, a repumper is required on the 4d\,$^2$D$_{3/2}\rightarrow5$p\,$^2$P$_{1/2}$ transition at 1092~nm. Also, once the ion is excited into the long-lived 4d\,$^2$D$_{5/2}$ level, a rapid means to return the ion from the upper state is necessary to reduce dead time in the laser cooling\,/\,probe\,/\,detection cycle. The method of choice is to use a state clearout source at 1033~nm to excite the 4d\,$^2$D$_{5/2}\rightarrow5$p\,$^2$P$_{3/2}$  transition whereupon rapid decay at the ns time scale returns the ion to the ground state.

%\section{Setup}

%\begin{figure}
%{\includegraphics[width=0.45\columnwidth]{YbCrossSections3}}
%{\includegraphics[width=0.45\columnwidth]{YbGain}}
%\caption{(a) Cross-sections for Yb-doped fiber. (b) Calculated gain \cite{Pask1995a} for a 1\;m (blue) and a 10\;m fiber (black). The pump wavelength is 980\;nm, %and the curves are labeled with the launched pump power in mW.}
%\label{crossSections}
%\end{figure}

\begin{figure}
{\includegraphics[width=1.0\columnwidth]{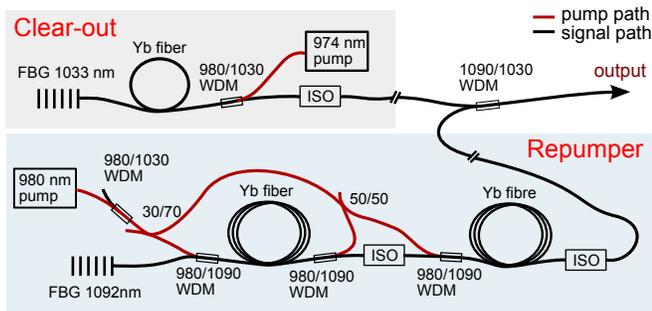}}
\caption{Amplified spontaneous emission source layout. WDM: wavelength division multiplexer; FBG: fiber bragg grating; ISO: optical isolator.} %The output from the source was taken from port A in the experiments described in this work}
\label{setup}
\end{figure}
\begin{figure}
{\includegraphics[width=0.85\columnwidth]{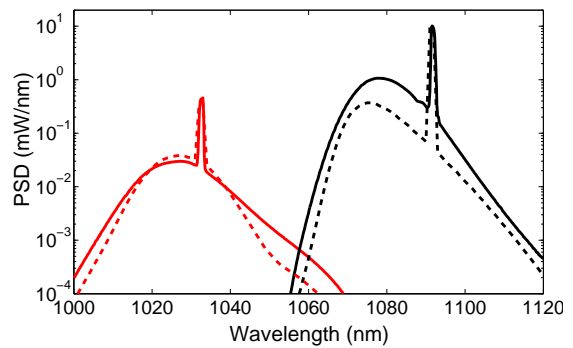}}
\caption{Measured clearout (solid red) and repumper (solid black) output power spectral densities (PSD). %Spectra of the repumper (black) and clear-out (blue) sources at output port A. 
 Also shown are spectra based on a numerical model (dashed).}%for the repumper (gray) and clear-out (gray dashed). }
\label{PSD}
\end{figure}

The clearout and repumper are based on ASE in Yb-doped fibers.
% the relevant cross sections of which  are shown in Fig.\;\ref{crossSections}a.
% Below about 1050\;nm, Yb-doped fiber behaves as a 'three level' system, whereas 'four level' behavior is seen at longer wavelengths, which means that as the fiber length is increased, the peak gain moves to longer wavelengths.
% (Fig.\;\ref{crossSections}b).
%
%This behavior can easily be deduced from the absorption and emission cross section in Fig.\;\ref{crossSections}a: a
% Fig.\ref{ the relevant cross-sections are shown in  
%
The layout is shown in Fig.\;\ref{setup}. The main part of both sources is an Yb-doped fiber, the back end of which is spliced to a narrowband (1-nm) fiber Bragg grating (FBG) reflector. The other (output) end is left open. For the clearout radiation at 1033~nm, a backward pumped 80-cm Yb-doped fiber  is used together with a two-stage, polarization insensitive isolator that prevents lasing due to optical feedback. The output spectrum is shown in Fig.\;\ref{PSD} (solid red line) at an approximate maximum pump power of 80\;mW. 

The power requirement for the repumper is three orders of magnitude higher than for the clearout, and this, together with the reduced gain at the longer wavelength, complicates the design. In the first stage, the repumper uses a 10-m Yb-doped fiber  that is pumped at both ends. To boost the output power and filter the spectrum, another 10-m Yb-doped fiber is used. With only forward pumping, this second fiber amplifies the signal at 1092\;nm while at the same time attenuating heavily radiation at 1033\;nm, which is important to avoid clearing out the clock state. The resulting output spectrum is shown in Fig.\;\ref{PSD} (solid black) for a total pump power of roughly 650\;mW. 

To guide the design, a simple numerical two-level model \cite{Desurvire2002a} was developed that takes into account the gain medium, the FBGs, the WDMs and losses in the isolators. Without detailed information on the doping profile, a step-like profile was assumed, the width of which was adjusted so that the peak experimental and numerical PSDs matched approximately. %he correct, fiber specific pump absorption.
The output of this simple model for the repumper and clearout is shown by the black and red dashed curves, respectively. The higher-than-expected background in the repumper spectrum is probably caused by reflections from optical components unaccounted for in the model or from a shortcoming of the two-level model used.

%The dashed curve shows spectrum when a constant 'background' reflectivity of only 0.2\% is included in the first stage of the repumper. Thus, for a low background, spurious reflections must be minimized. 

%The required pump power could be further reduced by shortening the fiber, the optimum being around XX cm (check this with Yb 800 fiber!).; however, the power requirements for reliable clear out operations are low enough so that further optimization was not necessary.
%The corresponding spectrum from a numerical model is shown by the gray curve. Again, to roughly match the peak PSD, the pump 
%power level was adjusted by XX\%.    

The ASE device was tested and compared against 1092-nm and 1033-nm lasers used in the Sr$^+$ single-ion optical clock facility at the National Research Council of Canada \cite{Dube2013a, Dube2014a}. Briefly, a single atomic ion of $^{88}$Sr$^{+}$ is held in an electrodynamic trapping field using a miniature radio frequency endcap trap~\cite{Dube2013a}.  The ion trap is operated at a frequency of $\Omega$~=~2$\pi\times$14.4~MHz with a voltage amplitude of $V_o$ = 215~V. A frequency stabilized, direct diode laser source at 422~nm is used as the primary laser cooling\,/\,fluorescence source. A diode-pumped fiber laser usually provides the repump for the 4d\,$^2$D$_{3/2}\rightarrow5$p\,$^2$P$_{1/2}$ transition at 1092~nm while state clearout is achieved with an external-cavity diode laser at 1033~nm. Both the 1092-nm and 1033-nm laser sources are frequency stabilized to a reference Fabry-Perot cavity whose absolute length is controlled by reference to a 633~nm polarization-stabilized HeNe laser.  Detected fluorescence rates of 10$^4$ counts per second are observed with a signal to background ratio of better than 50.  Interruption of this strong S-P fluorescence corresponds to probe excitation into the upper state of the clock transition.  %The probe system has been described in greater detail elsewhere~\cite{Dube2009a} and consistently shows resolution of the ion reference transition equivalent to the Fourier transform of the pulse width down to line resolutions of 4\;Hz.  
 The probe system~\cite{Dube2009a} consistently shows line resolutions down to 4\;Hz. A removable mirror %was inserted into the beam combining optics and 
 was used to rapidly switch between the  laser and ASE
  light sources. By enlarging the ASE beams as much as possible before beam combination, beam waists at the ion of approximately 41\;\textmugreek m and 54\;\textmugreek m were obtained for the repumper and clearout, respectively, with 64\% losses between source output and the ion.

%\section{Scattering rates}

\begin{figure}
{\includegraphics[width=0.85\columnwidth]{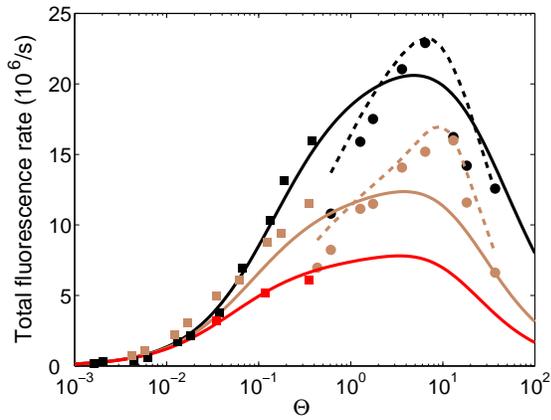}}

\caption{Total fluorescence rates for the ASE and laser repumpers as a function of the effective Rabi frequency squared per bandwidth $\Theta$. The solid circles represent data obtained with the laser repumper while the solid squares show data using the ASE repumper. Three cooling laser intensities were used: 550\;W/m$^2$ (red), 1100\;W/m$^2$ (brown) and 3300\;W/m$^2$ (black), corresponding to the Rabi frequencies $\Omega_\mathrm{c}/\Gamma$ = 1.2, 1.7, and 2.9, respectively. Theoretical values are shown with solid (ASE) and dashed (laser) lines. }

%The repumper intensity used in the theoretical calculations was 2.87 MW/m$^2$, a value extracted from the light shifts (Fig. 5).}
\label{scatteringRates}
\end{figure}

\begin{figure}
{\includegraphics[width=0.85\columnwidth]{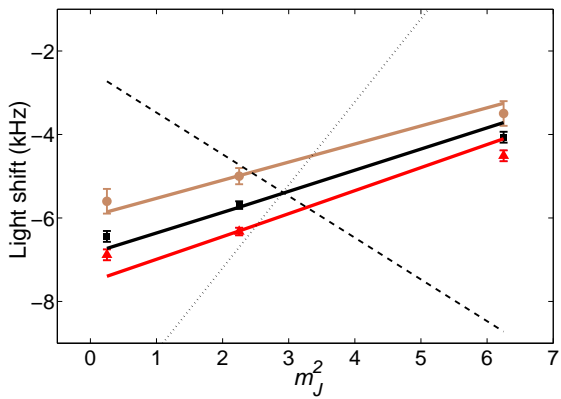}}
%{\includegraphics[width=0.45\columnwidth]{clearoutLightShifts2}}
\caption{Measured light shifts of the Zeeman components caused by the ASE repumper as a function of m$^2_J$ of the upper state of the clock transition. Solid lines show theoretical predictions for an unpolarized ASE light source with the intensities 2.49\;MW/m$^2$ (brown), 2.87\;MW/m$^2$ (black), and 3.15\;MW/m$^2$ (red). Also shown is the expected behavior for $\pi$-polarized (dotted) and $\sigma$-polarized light (dashed).}
\label{lightShifts}
\end{figure}

For clock operation, the cooling light scattered by the ion must be sufficient for reliable state detection. The measured total ion fluorescence rates are shown in Fig.\;\ref{scatteringRates} as a function of the effective Rabi frequency squared (proportional to intensity) per bandwidth $\Theta$ \cite{Lindvall2013a}. The total ion fluorescence rate was determined from that detected by carefully taking into account the detector response and losses in the imaging system (geometry and lens coatings). The power of a laser repumper is obtained from the relation $P_{\mathrm{l}}=612.1\;\mathrm{W/m^2}\cdot w_\mathrm{l}^{-2}\cdot\Theta_{\mathrm{l}}$, where $w_\mathrm{l}$ is the beam waist. For a laser $\Theta_{\mathrm{l}} = (\Omega_\mathrm{r}/\Gamma)^2$, where $\Omega_\mathrm{r}$ is the Rabi frequency and $\Gamma/2\pi=21.6$\;MHz is the natural linewidth of the P$_{1/2}$ excited state. For the broadband ASE repumper, the power spectral density (PSD) at resonance is given by $\Phi_{\mathrm{ASE}}$=5.415\;\textmugreek W/nm$\cdot(w_{\mathrm{ASE}}/\lambda)^2\cdot\Theta_{\mathrm{ASE}}$, where $\lambda$ is the wavelength and %$\omega_{\mathrm{ASE}}=41$\;\textmugreek m the beam waist. In this case
 $\Theta_{\mathrm{ASE}}=\Omega_\mathrm{r}^2/(b \Gamma)$, where $b$ is the (Lorentzian) linewidth of the source. For the theoretical results shown in Fig.\;\ref{scatteringRates}, the median of the intensities from light shift measurements reported in Fig.\;\ref{lightShifts} was used to relate the power of the ASE source to the intensities. With no free parameters, the agreement between theory and experiment is quite good.
 
%The inset in Fig.\;\ref{scatteringRates} shows the scattering rate obtained with the ASE repumper at full power as the cooling laser is tuned towards resonance. With a low cooling intensity of 110\;W/m$^2$, the half width at half maximum (HWHM) is 14\;MHz, which is close to the natural HWHM of 10.7\;MHz. % and shows that the power broadening caused by the ASE repumper is insignificant (theoretically XX kHz). Also here, the agreement between theory and experiment is very good without any free parameters.

%The inset in Fig.\;\ref{scatteringRates} shows the scattering rate as a function of cooling-laser detuning for a low cooling intensity of 110\;W/m$^2$ %($\Omega_c/\Gamma = 0.53$) and full ASE power (red curve). The agreement with theory (dashed blue curve) is good and the half width at half maximum (HWHM) %of 14\;MHz is close to the natural HWHM of 10.8\;MHz.

%\section{Ion temperatures}

\begin{figure}
{\includegraphics[width=0.85\columnwidth]{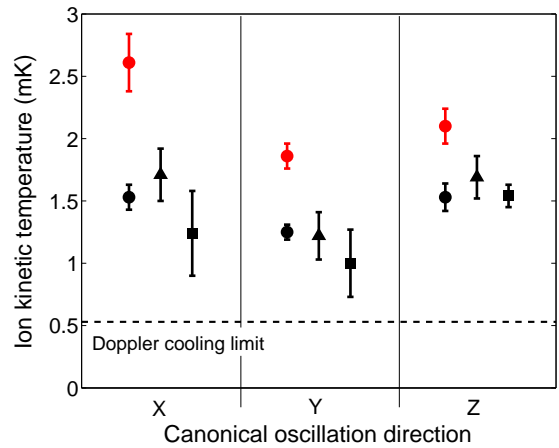}}
%{\includegraphics[width=0.49\columnwidth]{clearoutModulation}}
\caption{Measured ion kinetic temperatures using Zeeman-resolved motional sidebands with a laser repumper (red circles) and the ASE repumper (black circles). The 422~nm laser cooling intensity was 1700~\;W/m$^2$. Temperature measurements with a collapsed Zeeman spectrum (low B field) gave similar results (ASE source only, black triangles). Also shown is the measured temperature when the cooling intensity was lowered to 930\;W/m$^2$ (ASE source only, black squares). The errorbars equal one standard deviation. The Z-axis is the symmetry axis of the endcap trap.}
\label{temperature}
\end{figure}

%When displaced from the trap center, an ion experiences the trapping RF field, which causes micromotion induced second order Doppler and Stark shifts; 
%Once the black-body radiation shift uncertainty is reduced to the $10^{-18}$ level, the second-order Doppler shift due to the thermal motion of the ion is one of the dominant uncertainty mechanisms; therefore, it is important to characterize the ion temperatures when a broadband, high-power ASE repumper is used. 

In order to minimize the second-order Doppler shift due to thermal motion and its uncertainty, the ion temperature must be characterized. Temperature was determined by measuring the ion thermal velocity using ratios of the secular sidebands to the carrier \cite{Dube2013a}. With resolved Zeeman sidebands ($B=0.2\;\mu$T), the kinetic temperatures in the three canonical trap directions are shown in Fig.\;\ref{temperature} for the laser and ASE repumpers (red and black circles, respectively). The cooling laser intensity was 1700\;W/m$^2$. For the ASE repumper, the temperature was also determined using a collapsed Zeeman spectrum (low B-field), the result of which (black triangles) was in very good agreement with the previous result. By lowering the cooling intensity to 930\;W/m$^2$, a reduction in temperature could be observed (black squares). While both sources enable cooling close to the Doppler limit, the ASE repumper seems to yield somewhat lower temperatures.

%\section{Light shifts}

Due to the broad linewidths, the power needed for efficient repumping and clearout is about four orders of magnitude larger for the ASE sources than for lasers; therefore, the light shifts must be measured so that sufficient extinction can be applied during the probe pulse. %The uncertainty budget should be unaffected at the 10$^{-18}$ level. 
 Fig.\;\ref{lightShifts} shows three measurements of the light shift caused by the ASE repumper. Like the  electric quadrupole shift \cite{Dube2005a}, the light shift depends linearly on $m_j^2$ of the 4d\,$^2$D$_{5/2}$ clock state. The data is well aligned with the theoretical predictions for a completely unpolarized repumper (solid lines). The light shifts were calculated by summing the contributions from all transitions listed in \cite{JiangD2009a} (tables 1--2), taking into account the fractions of $\pi$ and $\sigma$ polarization. The spectrum of the ASE source (Fig.\;\ref{PSD}) was accounted for by expressing it as a sum of Lorentzian lines with the same total area, and adding the light shifts caused by the individual Lorentzians. The only free parameter is the intensity, and a fit to the data yields 2.49\;MW/m$^2$, 2.87\;MW/m$^2$, and 3.15\;MW/m$^2$. These values agree with the experimental estimate of 3.9\;MW/m$^2$ for a perfectly aligned beam with a 1.0 Strehl ratio. Also shown is the expected behavior for $\pi$- (dotted) and $\sigma$-polarized light (dashed). The difference in slopes between the experiment and theory is probably due to the light not being perfectly unpolarized.

%Mention kHz/(mW/nm) or something...

%The light shifts are dominated by the contribution from the nearest 1033-nm 4d$^2$D$_{5/2}$-5p$^2$P$_{3/2}$ transition, with 10--20\% contributions from the 422-nm 5s$^2$S$_{1/2}$-5p$^2$P$_{1/2}$ and 408-nm 5s$^2$S$_{1/2}$-5p$^2$P$_{3/2}$ transitions. Corrections from the shorter-wavelength transitions are very small.
Light shifts caused by the clearout source were investigated by running it continuously at power levels where the clearout efficiency was very low. Extrapolating the data for the $m_J=-\frac{1}{2}\rightarrow\frac{1}{2}$ Zeeman component to 'working conditions' points towards 10\;Hz order of magnitude light shifts if the source is on continuously.

%The light shifts of the $m_J=-\frac{1}{2}\rightarrow\frac{1}{2}$ Zeeman component were investigated by running the clear-out continuously at power levels where the clear-out efficiency was very low. The data showed that extrapolation to 'working conditions' gives light shifts of the order of 10\;Hz if the 1033~nm source would be on continuously. 

%\section{Modulation}
%\begin{figure}
%{\includegraphics[width=0.45\columnwidth]{RepumperModulation}}
%{\includegraphics[width=0.45\columnwidth]{clearoutModulation}}
%\caption{Normalized power time series (black) of the pump-modulated ASE repumper (a) and clear-out (b). Dashed lines indicate expected power decay based on %the excited state lifetime. Also show in (b) is the clear-out efficiency for three different intensities.}
%\label{modulation}
%\end{figure}
A very convenient feature of the ASE light sources is that they can be electronically turned on and off during the clock cycle by modulating the pump diodes. %For clock operation, 
The important parameters are then the turn-on and turn-off times and, especially, the extinction ratio. %Fig.\;\ref{modulation} (black traces) shows the modulation response when the pump diodes for the repumper (a) and clear-out (b) are square-wave modulated at 20\;Hz and 40\;Hz, respectively. The large dynamic range with sub-ms temporal resolution was achieved with a biased photodiode and a HP3458A digital multimeter. The turn on times are a few millisecond. At turn off, three time constants can clearly be seen. When the power is high, stimulated emission dominates and the power drops quickly by a few orders of magnitude. After that, the decay reflects the upper state lifetime (840\;\textmugreek s) (dashed blue traces). At the $10^{-6}$ level, close to the noise floor, another, much larger time constant starts to dominate. This is a measurement artifact, since the same  behavior is also seen with the pump lasers only; therefore, the dashed blue lines should indicate the real time dependence at the lowest light levels.
%At low output powers, the light decay is exponential 
 %Soon after turn-off, the light decay is exponential with a time constant that reflects the Yb upper state lifetime ($\approx900$\;\textmugreek s). %Consequently, 
The turn-on time is about 4\;ms (0-90\%). At turn off, stimulated emission dominates and the power drops quickly by two orders of magnitude. After that, the light decay reflects the upper state lifetime ($\approx900$\;\textmugreek s). 
%the light shift should be at the $10^{-17}$ level at the beginning of the probe pulse if the average shift of the clock transition during a 100\;ms probe pulse is to be at the $10^{-19}$ level. 
 If the average fractional light shift of the clock transition during a 100\;ms probe pulse is required to be at the $10^{-19}$ level, then, due to the exponential decay of the ASE sources, the shift should be at the $10^{-17}$ level at the beginning of the probe pulse. The measured light shifts translate into an extinction ratio requirement at the start of the clock pulse of $\gtrsim10^{6}$ for the repumper and $ \gtrsim10^{4}$ for the clearout. Measured values indicate that the repumper  and clearout will decay to this level  after the current is turned off in approximately 8\;ms and 4\;ms, respectively. 

Clearout efficiency was tested with 5\;ms pulses by varying the output power of the ASE source. The efficiency was 100\% at medium to high output powers. A drop to about 90\% at an estimated 170\;$\frac{\mathrm{W}}{\mathrm{nm}}$/m$^2$ (or 
 2300\;W/m$^2$) was observed, corresponding to a light source output of only 2.2\;\textmugreek W/nm. % When the power was lowered to XX\% of the peak power, power levels of  The output power was gradually lowered while efficiency was recorded. 
%  Blue solid lines in Fig.\;\ref{modulation}b show the calculated probability of the electron remaining in the clock state when a 5-ms clear-out pulse with 16000\;$\frac{\mathrm{W}}{\mathrm{nm}}$/m$^2$, 650\;$\frac{\mathrm{W}}{\mathrm{nm}}$/m$^2$ and 160\;$\frac{\mathrm{W}}{\mathrm{nm}}$/m$^2$ is applied. 
Independent calculations showed that at an intensity level of 160\;$\frac{\mathrm{W}}{\mathrm{nm}}$/m$^2$, the probability of the electron remaining in the D-state has risen to approximately 10\%, in good agreement with what was experimentally observed.    
   
%\section{Conclusions}

%In conclusion, key characteristics of the novel ASE-based light sources have been presented and their suitability for use in state-of-the-art optical single-ion clocks has been verified. Further improvements in terms of lower light shifts could be  achieved using more narrowband FBGs and/or external narrowband interference filters. The developed numerical model can be used to further optimize the design, and if larger beam diameters have to be used or severe beam delivery losses need to be tolerated, additional gain stages can be added at very low cost. Similar repumper light sources could be made for Ca$^+$ (866 nm) and Ba$^+$ (650\;nm) using semiconductor gain media.

In conclusion, key characteristics of the ASE light sources have been presented and their suitability for use in state-of-the-art optical single-ion clocks has been verified. Still lower light shifts can be obtained using more narrowband FBGs and/or external narrowband interference filters. The numerical model can be used to further optimize the design, and if higher PSDs are needed additional gain stages can be easily added. Similar repumpers could be made for Ca$^+$ (866 nm) and Ba$^+$ (650\;nm) using semiconductor gain media.

%\section{Acknowledgements}

This work was supported by the Academy of Finland (project 138894) and by the European Metrology Research Program (EMRP) in project SIB04. The EMRP is jointly funded by the EMRP participating countries within EURAMET and the European Union. TF acknowledges financial support from the European Commission (Marie Curie Integration Grant PCIG10-GA-2011-304084).
%\begin{thebibliography}{99}
%% Do not include separate BibTeX files; if BibTeX is used,
%% paste the output (contents of .bbl file) here.
%\bibitem{revtex-au} \url{https://authors.aps.org/revtex4/}.
%\bibitem{osastyle} \url{http://www.opticsinfobase.org/submit/style/jrnls_style.cfm}.
%\end{thebibliography}

%\bibliographystyle{ol}
%\bibliography{IonClock}


%\bibliographystyle{}
%\bibliography{IonClock}
%\printbibliography
\end{document}